\documentclass[sn-mathphys,Numbered]{sn-jnl}% %\documentclass{article}
\usepackage[utf8]{inputenc}
\usepackage{graphicx}%
\usepackage{multirow}%
\usepackage{amsmath,amssymb,amsfonts}%
\usepackage{amsthm}%
\usepackage{mathrsfs}%
\usepackage[title]{appendix}%
\usepackage{xcolor}%
\usepackage{textcomp}%
\usepackage{manyfoot}%
\usepackage{booktabs}%
\usepackage{algorithm}%
\usepackage{algorithmicx}%
\usepackage{algpseudocode}%
\usepackage{listings}%

\usepackage{ulem}
\usepackage{xspace}
\usepackage{comment}
\usepackage{aas_macros}

\newcommand{\ben}{\begin{enumerate}}
\newcommand{\een}{\end{enumerate}}
\newcommand{\Fermi}{\textit{Fermi}\xspace}

\newcommand{\gray}{$\gamma$-ray\xspace}
\newcommand{\grays}{$\gamma$ rays\xspace}

\renewcommand{\subsectionautorefname}{section}
\renewcommand{\subsubsectionautorefname}{section}
\newcommand{\Autoref}[1]{%
  \begingroup%
  \def\chapterautorefname{Chapter}%
  \def\sectionautorefname{Section}%
  \def\subsectionautorefname{Section}%
  \def\subsubsectionautorefname{Section}%
  \autoref{#1}%
  \endgroup%
}

\begin{document}

\title[Diffuse]{
Diffuse continuum emission and large extended sources at MeV energies}

\author*[1]{\fnm{Markus} \sur{Ackermann}}\email{markus.ackermann@desy.de}

\author*[2]{\fnm{Denys} \sur{Malyshev}}\email{denys.malyshev@astro.uni-tuebingen.de}
\author*[3]{\fnm{Dmitry V.} \sur{Malyshev}}\email{dmitry.malyshev@fau.de}

\affil[1]{\small Deutsches Elektronen-Synchrotron DESY, Platanenallee 6, 15738 Zeuthen, Germany}

\affil[2]{\small Institut für Astronomie und Astrophysik Tübingen, Universität Tübingen, Sand 1, 72076 Tübingen, Germany}
\affil[3]{\small Erlangen Centre for Astroparticle Physics, Nikolaus-Fiebiger-Str.\ 2, 91058 Erlangen, Germany}

\abstract{
Future \gray survey instruments, such as \textit{newASTROGAM} and \textit{AMEGO-X}, will significantly improve previous and current all-sky surveys at MeV energies.
In this paper we discuss the continuum emission from the Milky Way, two prominent large extended sources, the \Fermi bubbles and Loop I, and the extragalactic \gray background.
We highlight the importance of measurements in the MeV to GeV energy range for understanding CR production and propagation in the Galaxy, for the determination of the nature of the \Fermi bubbles and Loop I, and for exploring the origin of the extragalactic \gray background. }


\maketitle

\section{Introduction}
\label{sec:intro} 
Diffuse \gray emission arises from various origins and on vastly different scales. It dominates the total integrated \gray flux in the MeV and GeV range.
In contrast to line emission from positron annihilation \cite{1973ApJ...184..103J,1978ApJ...225L..11L,2008Natur.451..159W,2020ApJ...897...45S} and the decay of radioisotopes, such as $^{26}$Al~\cite{1995A&A...298..445D,2015ApJ...801..142B} and $^{60}$Fe \cite{2007A&A...469.1005W,2020ApJ...889..169W}, which are also important sources of diffuse emission in our Galaxy, diffuse continuum emission arises from the interactions of non-thermal particles with radiation and matter. 
We focus on this continuum emission below rather than the line emission, which is discussed elsewhere in this edition. %interchapter references?.

The diffuse \gray emission has its origins in the solar system, the Milky Way, and the Universe beyond. 
Within the solar system, the most prominent sources of diffuse radiation are cosmic-ray (CR) interactions with the Earth atmosphere (e.g.,~\cite{1972ApJ...177..341K,2009PhRvD..80l2004A}) and the radiation field of the sun~\cite{2008A&A...480..847O,2011ApJ...734..116A}. 
The Galactic diffuse emission is produced by the interactions of CRs with the interstellar gas (ISG) and interstellar radiation fields (ISRF). 
It is customary to distinguish the bulk diffuse Galactic emission (DGE) that is fueled by quasi-continuous CR production, propagtion and escape on the scale of the entire Milky Way from individual, large-scale structures that arise from transient and/or localized CR injection events (e.g., Loop I and Fermi Bubbles, see below). 

The extragalactic diffuse emission, also called the extragalactic \gray background (EGB) is dominantly produced by faint or distant \gray sources that are unresolved by current instruments. 
Known populations of extragalactic \gray sources include active galactic nuclei (AGN)~\cite{2008ApJ...672L...5I}, star-forming galaxies~\cite{2014ApJ...786...40L}, but also transients such as supernovae (SN), dominated by SN Type Ia (SNIa)~\cite{1975ApJ...198..241C,1999ApJ...516..285W,2016ApJ...820..142R}, and gamma-ray bursts (GRB)~\cite{2019ApJ...878...52A}. 
Other potential contributions to the EGB arise from CR interactions with the cosmic microwave background (CMB)~\cite{1975Ap&SS..32..461B} and beyond-the-standard-model (BSM) physics processes~\cite{2001PhRvL..87y1301B,2021RPPh...84k6902C}. 
Due to its predominant origin from unresolved sources, the measured diffuse extragalactic emission depends on the sensitivity and angular resolution of the instrument used. 
We therefore use a convention here introduced in~\cite{2015ApJ...799...86A} to distinguish between the isotropic \gray background (IGRB) that encompasses only the diffuse extragalactic emission and the \textit{total EGB} that includes the IGRB and all resolved extragalactic sources. 
The latter is independent of the instrument used to measure it and encompasses the entire extragalactic \gray emission from astrophysical sources. 

This manuscript summarizes the current knowledge and open science questions related to the diffuse \gray emission in the MeV range, focusing on the Galactic and extragalactic components.
\Autoref{sec:galactic} discusses the DGE. 
\Autoref{sec:extended} focuses on the largest extended features of the Galactic \gray sky, the \Fermi bubbles and Loop I. 
\Autoref{sec:extragalactic} discusses the extragalactic diffuse background, what is known about its origin, and the prospects for future missions to constrain the contributions of various source populations and new physics.
%\dima{Section should start with capital S.} %Done

\section{Diffuse Galactic emission}
\label{sec:galactic}
%Past measurements, Expectations of the MeV emission from CR interactions in the ISM, Circumgalactic emission, Nearby galaxies (LMC, SMC, Andromeda) Potential of future measurements.

The DGE arises from the interactions of CRs with the ISG and the ISRF via various processes, such as bremsstrahlung, inverse Compton (IC) scattering, and pion decay. 
Observations of the DGE play a crucial role in understanding CR propagation in the Milky Way and the interstellar medium~(ISM) properties of our Galaxy. 
It has been observed by several generations of \gray telescopes, including the \textit{OSO-3}~\cite{1972ApJ...177..341K}, \textit{SAS-2}~\cite{1975ApJ...198..163F}, \textit{COS-B}~\cite{1975SSI.....1..245B}, \textit{EGRET}~\cite{1997ApJ...481..205H} and the \Fermi-LAT~\cite{2012ApJ...750....3A} instruments at energies above few tens of MeV. 
\textit{COMPTEL}~\cite{1994A&A...292...82S}, \textit{OSSE}~\cite{1997ApJ...483L..95S} \textit{INTEGRAL/SPI}~\cite{2008ApJ...679.1315B} and, recently, \textit{COSI}~\cite{2023ApJ...959...90K} provide measurements of the DGE in the energy range from 20 keV to few tens of MeV.  

\begin{figure}[t]
\centering
\includegraphics[width=0.8\textwidth]{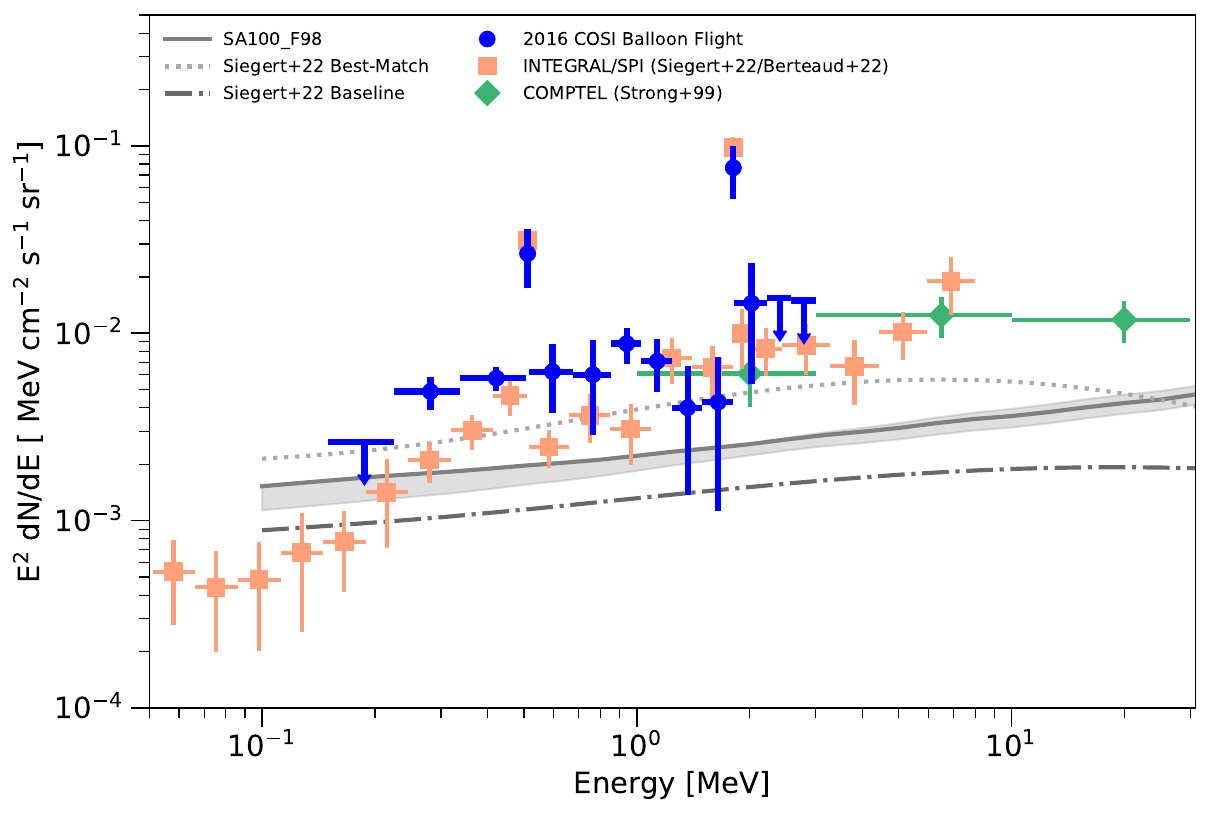}
\caption{Measurements of the DGE in the direction of the inner Galaxy during the 2016 \textit{COSI} balloon flight~\cite{2023ApJ...959...90K}, by COMPTEL~\cite{1994A&A...292...82S,1999ApL&C..39..209S}, and SPI~\cite{2022A&A...660A.130S,2022PhRvD.106b3030B} in comparison to expectations~\cite{2022ApJS..262...30P,2022A&A...660A.130S}. The models do not include line emission visible in the data at 511~keV (e$^{+}$/e$^{-}$ annihilation) and 1.81~MeV ($^{26}$Al decay). Figure adapted from~\cite{2023ApJ...959...90K}.}
\label{fig:karwin}
\end{figure}

\autoref{fig:karwin} shows the DGE in the direction of the inner Galaxy as measured during the 2016 \textit{COSI} balloon flight, by the \textit{COMPTEL} instrument on the \textit{Compton Gamma-Ray Observatory (CGRO)}, and the \textit{Spectrometer on INTEGRAL (SPI)} in the energy range between 50 keV and 30 MeV in comparison to expectations.
A common method for studying the DGE in the MeV and GeV energy regimes involves modeling the CR production and propagation in the Milky Way, as well as the emission of \grays from CR interactions with the ISG and ISRF via a propagation code, such as \textsc{Galprop}~\cite{2011CoPhC.182.1156V}, \textsc{Dragon}~\cite{2017JCAP...02..015E}, or \textsc{Picard}~\cite{2014APh....55...37K}. 
Comparison of the model predictions with the observed DGE and local observations of primary and secondary CRs allow one to study and constrain the parameters that enter the model, such as the CR source distribution, diffusion coefficients, the roles of convection and/or re-acceleration of CRs in the interstellar medium, CR halo size (e.g.,~\cite{2012ApJ...750....3A}). 
In addition, the DGE can be used to study ISM properties, such as the ISG and ISRF distribution, and the Galactic magnetic field structure.
The excess of the observed continuum DGE over predictions from CR propagation models above few hundred keV that is visible in \autoref{fig:karwin} is generally attributed to contributions from unresolved \gray sources~\cite{2023ApJ...943...48T}. However, also other contributions are possible, such as a more intense radiation field in the Galactic bulge~\cite{2008ApJ...682..400P,2011ApJ...739...29B} or a contribution from the annihilation of dark matter particles in the MeV mass range~\cite{2025PhRvL.134j1001L}.

\begin{figure}[h!p]
\centering
\includegraphics[width=0.52\textwidth]{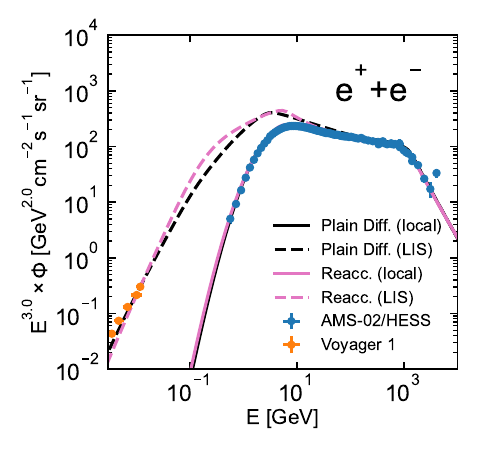}\\
\includegraphics[width=0.98\textwidth]{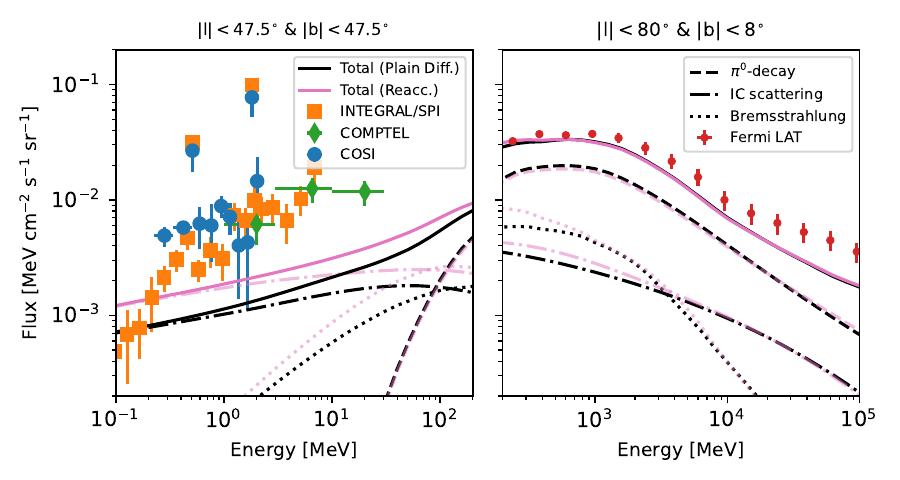} 
\caption{\textit{Upper panel:} Comparison of the locally observed primary CR electron + positron spectra by AMS-02~\cite{2014PhRvL.113v1102A} and HESS~\cite{2008PhRvL.101z1104A} and the spectrum observed by Voyager-1 in interstellar space~\cite{2016ApJ...831...18C} (denoted as ``LIS'' in panel legend) to expectations from two different models of CR propagation. Both models were computed with \textsc{Galprop}~\cite{2022ApJS..262...30P}, assuming plain diffusion + convection of CR in the ISM in one case (black), and diffusion with re-acceleration of CR in the ISM (pink) in the other case. \textit{Lower panel:} Expected DGE emission in the MeV and GeV bands from the two CR propagation models shown in the upper panel. The expected emission (black/pink solid lines) is compared to the MeV observations displayed in \autoref{fig:karwin}, and to the measurement of the diffuse emission by \Fermi LAT in~\cite{2012ApJ...750....3A}~(red bars, right panel). \Fermi-LAT measurements have been published for a different region of the sky than the MeV measurements in the left panel. The corresponding region is indicated above the respective panel. Individual contributions from various interaction processes are shown as dashed/dotted lines (see legend).}
\label{fig:galprop_models} 
\end{figure}

 \autoref{fig:galprop_models} demonstrates the importance of the MeV window for constraining CR propagation. The figure shows two different models of the Galactic diffuse emission computed with \textsc{Galprop}(v57)~\cite{2022ApJS..262...30P} in comparison to the MeV measurements of the DGE featured in \autoref{fig:karwin}, and measurements above 200~MeV obtained from the \Fermi-LAT data for a different region of the sky~\cite{2012ApJ...750....3A}. Both models use the same distribution of CR sources for injecting CR (Galactic Pulsar distribution,~\cite{2004A&A...422..545Y}), and are tuned to the same set of CR observations. For CR electrons+positrons, these are the low-energy ($<$~50~MeV) measurements by \textit{Voyager 1} in interstellar space~\cite{2016ApJ...831...18C}, and high-energy measurements by \textit{AMS-02}~\cite{2014PhRvL.113v1102A} and \textit{HESS}~\cite{2008PhRvL.101z1104A}, both performed at or near Earth. 

 The first model assumes plain CR diffusion with a break in the rigidity dependence of the diffusion coefficient at 4~GV. In addition, convection of CRs is modeled with a convection speed increasing with distance from the Galactic plane ($v_{0}=0$~km\,s$^{-1}$, $dv/dz = 10$~km\,s$^{-1}$\,kpc$^{-1}$). 
 Plain diffusion has been shown to describe the Galactic synchrotron emission spectrum in the radio band better than models which assume (strong) re-acceleration of CR in the interstellar medium~\cite{2011A&A...534A..54S}.
 The second model shown is one with moderate re-acceleration. Such models describe the light element (p, He) CR spectra best, according to an extensive study of propagation parameters in~\cite{2016ApJ...824...16J}. 
 Re-acceleration of CR in the ISM typically removes the necessity for a break in the rigidity dependence of the diffusion coefficient. Moderate re-acceleration also remains consistent with synchrotron observations~\cite{2018MNRAS.475.2724O}. \textsc{Galprop} configuration files for the two models shown in \autoref{fig:galprop_models} can be found in~\cite{ackermann_2025_15007321}.
 
Below few tens of MeV, the DGE is almost entirely produced from IC scattering of CR electrons and positrons with the ISRF. The strongest variations of the intensity of the DGE between the two propagation scenarios depicted in \autoref{fig:galprop_models} are visible in this energy range. At higher energies, interactions of CR nuclei with the ISG resulting in the production of $\pi^0 \rightarrow \gamma\gamma$ strongly dominate the DGE. 
Bremsstrahlung from the interactions of CR electrons with the ISG remains subdominant in the entire MeV energy range, but contributes significantly to the total DGE from few MeV to few hundreds of MeV.
The bremsstrahlung and pion-decay components are strongly correlated to the distribution of the ISG column density and, correspondingly, their intensity exhibits strong intensity variations over the sky on all scales. 
In contrast, the IC component shows smooth variations over the sky, determined by CR electron and radiation field densities along each line of sight (see, e.g.,~\cite{2019PhRvD..99d3007O}).
 The kinematics of the pion decay lead to a characteristic steep spectrum for the pion-decay component below $\frac{1}{2} m_{\pi^{0}} \approx 67.5$~MeV~\cite{1971NASSP.249.....S}, allowing to distinguish it from bremsstrahlung based on its spectral shape. 

 A precise measurement of the spectrum and spatial distribution of the DGE in the transition region between 10 MeV and 100 MeV is therefore crucial to disentangle the contributions from the different processes. 
 The CR energies that are responsible for the DGE in this energy range are typically around few tens of MeV to few GeV, a range where local measurements of CR spectra do not reflect their local interstellar spectrum (LIS) due to strong and time-dependent solar modulation effects~\cite{2013LRSP...10....3P}. 
 The Voyager measurements of LIS spectra are limited to $\lesssim50$~MeV for electrons and $\lesssim350$~MeV for protons~\cite{2016ApJ...831...18C}. 
 Precise \gray measurements of the DGE can indirectly probe the LIS CR spectrum above this energy range, and therefore, as discussed above, provide important constraints on CR propagation models. 
 This can also be seen in the upper panel of \autoref{fig:galprop_models} where the two models displayed exhibit clear differences in the electron + positron spectrum in the range between few tens of MeV and few GeV, where solar modulation effects are strong and no direct LIS measurements are available.
 In addition, the strong dominance of IC scattering and pion decay at low and high energies, respectively, unlocks the potential to improve our understanding of the ISRF and ISG distributions based on measurements of the spatial distributions of the \gray emission.

 The measurement itself, however, is challenging for various reasons. 
 The key energy range between 10 MeV and 100 MeV needs hybrid instruments that combine a Compton telescope with a pair detection instrument, such as the proposed \textit{newASTROGAM}~\cite{Berge:2025ICRC} and \textit{AMEGO-X}~\cite{2022JATIS...8d4003C} concepts. 
 At energies above 10~MeV the effective area of typical Compton telescopes decreases rapidly due to the decreasing Compton cross section and increasing leakage of Compton photons, while maintaining an $\mathcal{O}(1^{\circ})$ angular resolution~\cite{2022JATIS...8d4003C}. 
 The effective area of pair creation instruments increases with energy, but their angular resolution deteriorates quickly for energies below 100~MeV due to the increasing effects of multiple scattering of the electron-positron pairs in the detector material. 
 Their angular resolution reaches $\mathcal{O}(10^{\circ})$ below few tens of MeV~\cite{2018JHEAp..19....1D}.

However, good angular resolution is crucial for disentangling the DGE from the foreground of unresolved Galactic and background of extragalactic sources, as well as extended diffuse structures such as the \Fermi Bubbles and Loop~I discussed in \autoref{sec:extended}. 
The upcoming \textit{COSI} Compton telescope \cite{2023arXiv230812362T} will be an important first step forward. \textit{COSI} will allow to resolve fainter foreground sources with its unprecedented continuum sensitivity around 1~MeV in energy, yielding a higher accuracy estimate of the contribution of unresolved sources to the DGE. It will also improve the resolution of spatial variations in the continuum DGE in this energy range. A future hybrid Compton/Pair instrument with an energy range from sub-MeV to multiple GeV allows to study these sources and extended foregrounds over a wide energy range with good angular resolution and sensitivity below 10~MeV and above 100~MeV. These measurements will facilitate a high-confidence estimation of the foreground/background source contribution to the DGE also in the energy range between 10~MeV and 100~MeV, where it is hard to achieve both, good angular resolution and sensitivity, simultaneously.

Other important foregrounds and backgrounds comprise the IGRB (see also \autoref{sec:extragalactic}) and \grays from the Earth's atmosphere. 
The IGRB intensity can be distinguished from the DGE by its isotropic distribution over the sky, while the Earth's \gray emission is predominantly an issue in the Compton regime, since the Compton scattering angle can be large. 
Additional shielding by a thick active veto system or the rejection of large scattering angles can help to suppress the Earth's \gray background.

Exploring the spatial distribution of the DGE and correlations to ISRF and ISG requires an instrument with significantly higher effective areas in the MeV range than that of past and current generations of instruments. 
Around 100~MeV, the expected DGE intensity from pion decay and IC scattering towards the inner Galaxy is similar for both components, $\sim 2 \times 10^{-3}$~MeV~cm$^{-2}$~s$^{-1}$~sr$^{-1}$, according to the model shown in \autoref{fig:galprop_models}. 
However, at the Galactic poles the DGE intensity is more than an order of magnitude lower for IC scattering~\cite{2019PhRvD..99d3007O} and for the pion decay component~\cite{2012ApJ...750....3A}. 
Therefore, an instrument with an extended source sensitivity of $\lesssim 10^{-4}$~MeV~cm$^{-2}$~s$^{-1}$~sr$^{-1}$ in the energy range between 10~MeV and 100~MeV, such as \textit{newASTROGAM}~\cite{Berge:2025ICRC} or other proposed instruments, is required to study the DGE variations over the entire sky.

\section{\Fermi bubbles and Loop I}
\label{sec:extended}

\newcommand{\refGenFBs}{\cite{2010ApJ...724.1044S_Su_bubbles, 2012ApJ...753...61S, 2014ApJ...793...64A_FB_Fermi, 2017MNRAS.468.3051N, 2019A&A...625A.110H_LLB_Herold_Malyshev}\xspace}

\newcommand{\refGenLoopI}{\cite{1971A&A....14..252B, 2007ApJ...664..349W, 2015MNRAS.452..656V, 2016A&A...594A..25P, 2018Galax...6...62S, 2018Galax...6...56D, 2018Galax...6...27K, 2018Galax...6...26S, 2023CRPhy..23S...1L}\xspace}

\newcommand{\refFBmodels}{\cite{2011MNRAS.415L..21Z_Zobovas_burst_model, 2011ApJ731L17C_FB_bursts, 2011PhRvL.106j1102C_Crocker-Aharonian_FB_Model, 2012ApJ...756..181G_Guo_AGN, 2013MNRAS.436.2734Y_Karen_Yang, 
2015ApJ...808..107C, 2022NatAs...6..584Y}\xspace}

\newcommand{\refLoopImodels}{\cite{1971A&A....14..252B, 2007ApJ...664..349W, 2015MNRAS.452..656V, 2016A&A...594A..25P, 2018Galax...6...62S, 2018Galax...6...56D, 2018Galax...6...27K, 2018Galax...6...26S}\xspace}

\newcommand{\refLoopIXrayDist}{\cite{2013ApJ...779...57K, 2016A&A...595A.131L, 2016A&A...594A..78G, 2023CRPhy..23S...1L}}

\newcommand{\refLoopIMWDist}{\cite{2013Natur.493...66C, 2015MNRAS.452..656V, 2016A&A...594A..25P}}
% 1971A&A....14..252B - definitions of Loop I - Loop IV and first model as SNRs
% 2007ApJ...664..349W - Wolleben model of Loop I with colliding superbubbles

\newcommand{\refsFBsAGNs}{\cite{2011MNRAS.415L..21Z_Zobovas_burst_model, 2011ApJ731L17C_FB_bursts, 2012ApJ...756..181G_Guo_AGN, 2013MNRAS.436.2734Y_Karen_Yang, 2017MNRAS.467.3544S_Sarkar_OVIII-OVII, 2022NatAs...6..584Y, 2023ApJ...951...36S_Sarkar_minijets}\xspace}

\newcommand{\refsFBsSF}{\cite{2011PhRvL.106j1102C_Crocker-Aharonian_FB_Model, 2015ApJ...808..107C, 2017MNRAS.467.3544S_Sarkar_OVIII-OVII}\xspace}

\newcommand{\refsFBsFountain}{\cite{2011PhRvL.106j1102C_Crocker-Aharonian_FB_Model, 2015ApJ...808..107C, 2024ApJ...973...78S}\xspace}

\Fermi bubbles (FBs) 
\refGenFBs
%[ref: general FBs list] 
and Loop I
\refGenLoopI
%[ref: general Loop I list] 
are the largest extended sources in the \gray sky (apart from the Milky Way Galaxy itself and the isotropic \gray background).
In spite of many multi-wavelength observations of these objects and a significant modeling effort for FBs 
\refFBmodels
%[ref: FB models list] 
and Loop I 
\refLoopImodels
%[ref: Loop I models list], 
their origin is still under debate.
In particular, Loop I was originally discovered in radio observations~\cite{1971A&A....14..252B}.
Models based on the radio data include a nearby supernova remnant~\cite{1971A&A....14..252B} or superbubbles (supershells) created by stellar winds and supernovae of the local Scorpio-Centaurus OB association~\cite{2007ApJ...664..349W}.
The large size of Loop I on the sky is explained because the distance to the edges of the superbubble is comparable to the size of the superbubble itself.
One of the main arguments in favor of a local origin of Loop I is the polarization of light from stars behind a synchrotron emitting region~\cite{2015MNRAS.452..656V, 2018Galax...6...56D}.
A structure similar to the radio Loop I is also visible in the X-ray data, where it is often referred to as the North polar spur or, more recently, the \textit{eROSITA} bubbles~\cite{2020Natur.588..227P}.
The X-ray observations suggest, however, that the corresponding structure is located at a much larger distance than the local superbubble due to absorption of soft X-rays by the interstellar gas \refLoopIXrayDist.
Another argument in favor of the GC origin of Loop I is depolarization of low-frequency \textit{WMAP} and \textit{Planck} maps close to the Galactic plane \refLoopIMWDist.
%Thus X-ray observations favor an interpretation of Loop I as a giant outflow from the central region of our Galaxy.

The FBs were discovered \cite{2010ApJ...724.1044S_Su_bubbles}
in the \Fermi-LAT data after about a year and a half of operations of the \Fermi satellite~\cite{2009ApJ...697.1071A}. 
Their symmetrical distribution above and below the GC region suggests that the FBs are created by an activity in that region, such as a jet or an outflow from the supermassive black hole Sgr A*, the AGN scenario \refsFBsAGNs, a star-burst activity near the GC~\cite{2014MNRAS.444L..39L}, or regular star formation and supernovae explosions~\refsFBsSF.
In the AGN outflow or starburst scenarios the FBs are inflated on the timescale of millions to tens of millions of years.
In this case, the \gray emission is explained by inverse Compton scattering of high energy electrons with the ISRF.
Provided that the \gray emission from the FBs is observed above 100 GeV, the required energy of electrons is about 1 TeV~\cite{2014ApJ...793...64A_FB_Fermi}.
These electrons have a cooling time of 1 Myr or less~\cite{2014ApJ...793...64A_FB_Fermi}, which is much shorter than the timescale of formation of the FBs.
In order to explain the \gray emission from the FBs, a re-acceleration of electrons is necessary~\cite{2011PhRvL.107i1101M_Mertsch_Sarkar_FB_model}.
The FBs can also be modeled as a persistent structure in the Galaxy supported by the star formation and supernovae explosions near the GC.
In this case the \gray emission mechanism is typically attributed to interactions of hadronic CRs with gas, where the required density of CRs can be accumulated on timescales of hundreds of millions to billions of years~\refsFBsFountain.
Related features in microwave~\cite{2004ApJ...614..186F_WMAP_haze_Finkbeiner, 2012ApJ...755...69P_WMAP_haze, 2013A&A...554A.139P_Planck_haze} and radio~\cite{2013Natur.493...66C} wavelengths are explained due to the presence of (re-accelerated) secondary leptons~\cite{2015ApJ...808..107C}.
For recent reviews of the FBs and the \textit{eROSITA} bubbles/Loop I see~\cite{2023CRPhy..23S...1L, 2024A&ARv..32....1S_Sarkar_FB_review}.

\begin{figure}[h]%
\centering
\includegraphics[width=0.49\textwidth]{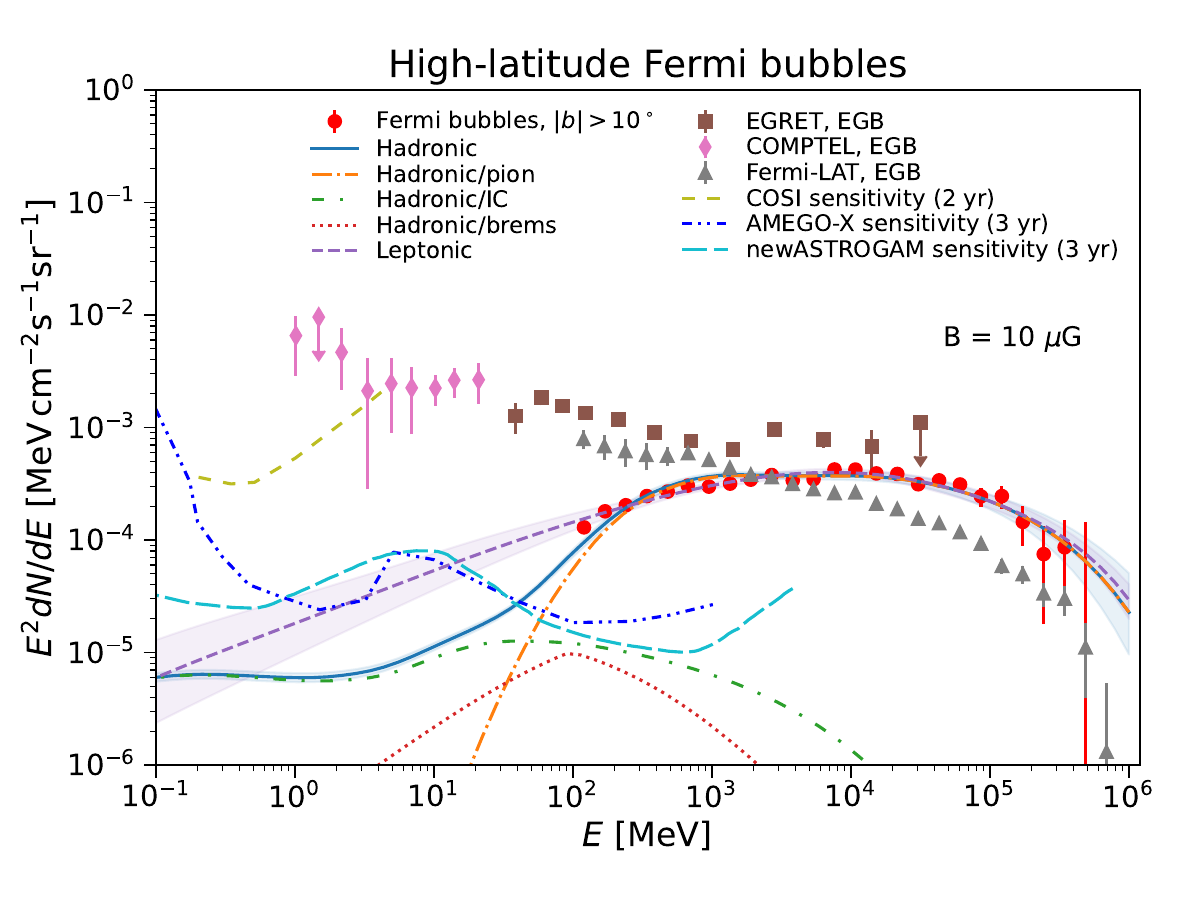}
\includegraphics[width=0.49\textwidth]{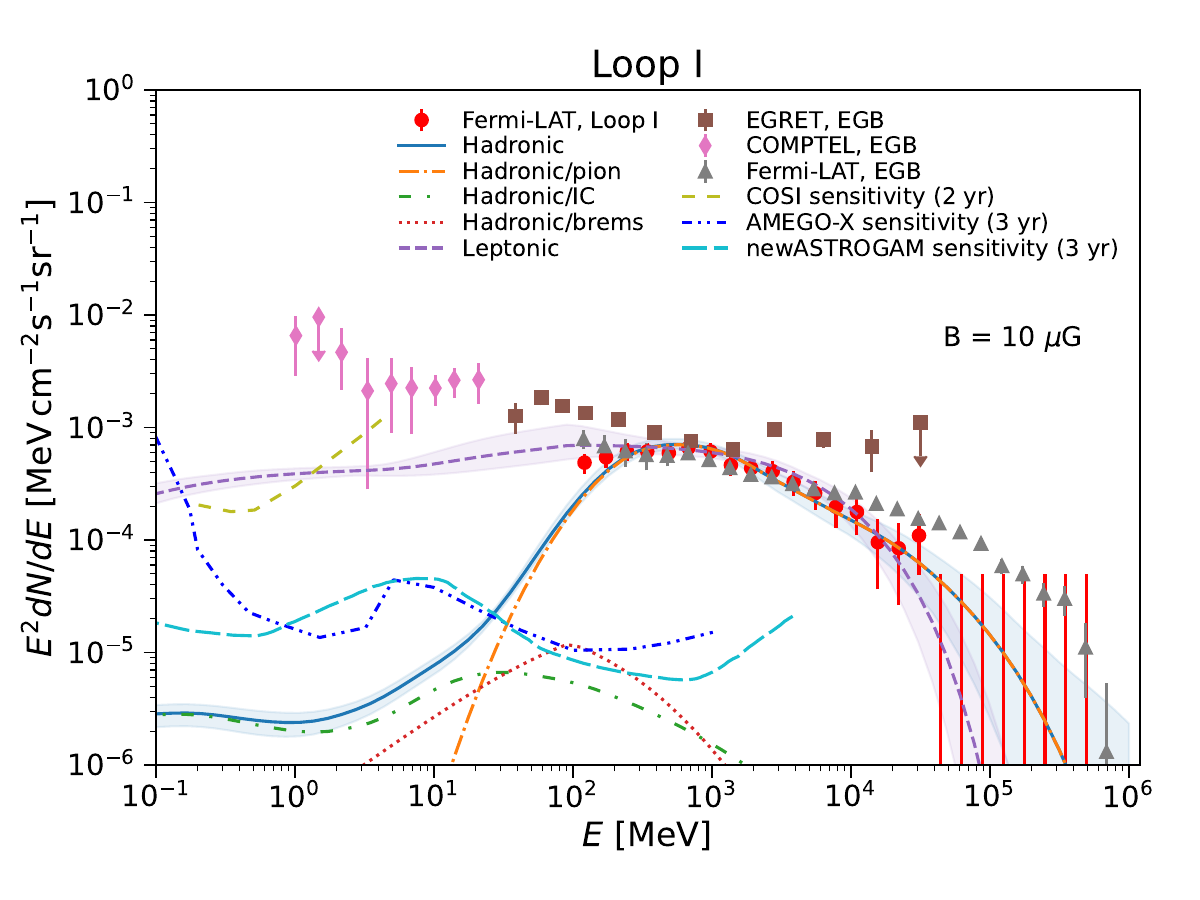}\\
\includegraphics[width=0.49\textwidth]{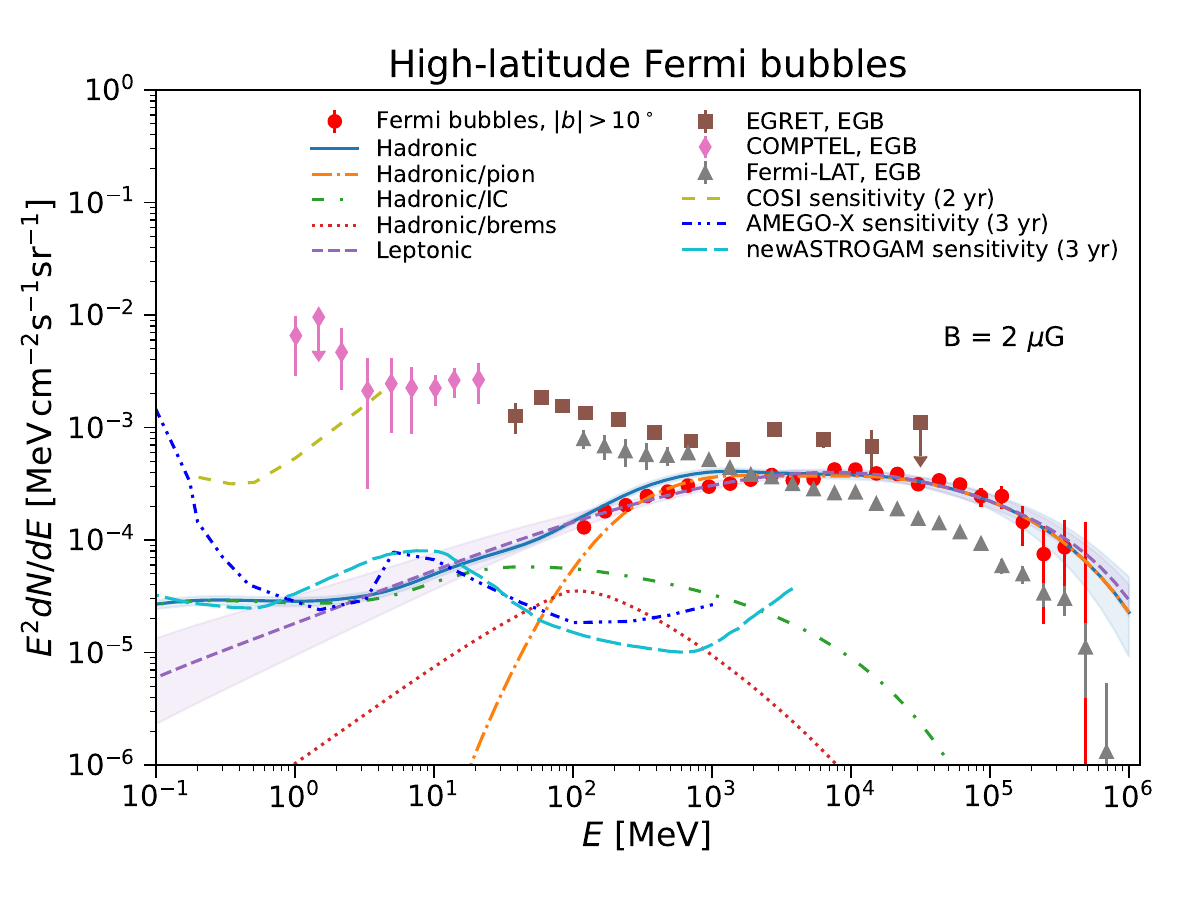}
\includegraphics[width=0.49\textwidth]{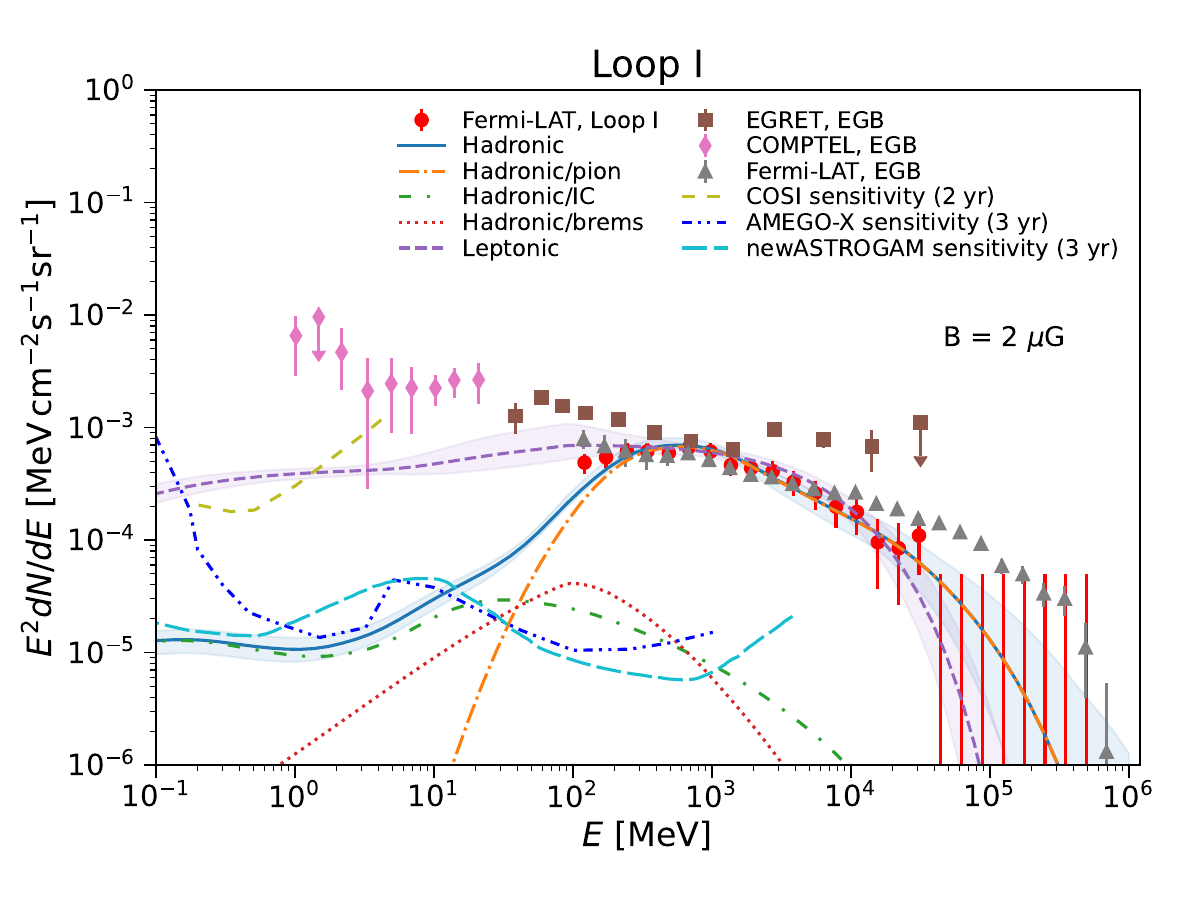}
\caption{Intensity of emission of the FBs (red circles, left panel) and Loop I  (red circles, right panel) at latitudes $|b| > 10^\circ$~\cite{2014ApJ...793...64A_FB_Fermi}.
Blue solid line shows the hadronic model of the \gray emission. Dashed orange line, green sparse dash-dotted line, and dotted red line show the primary $\pi^0$, secondary IC, and secondary bremsstrahlung components in the hadronic scenario respectively. Dashed purple line - leptonic scenario of \gray emission (dominated by IC emission). Bands show the 1 sigma model uncertainty ranges for statistical plus 10\% systematic uncertainties in the data. Pink diamonds, brown squares, and grey upward triangles show the extragalactic diffuse \gray background measured by 
\textit{COMPTEL}~\cite{2000AIPC..510..467W},
\textit{EGRET}~\cite{2004ApJ...613..956S}, and 
\Fermi~LAT~\cite{2015ApJ...799...86A} respectively.
Yellow sparse dashed lines show expected \textit{COSI} sensitivity after 2 years of observations~\cite{2023arXiv230812362T} for an extended source with the area of the 
high-latitude FBs $\Omega \approx 1$~sr
and Loop I $\Omega \approx 3$~sr respectively.
Expected \textit{AMEGO-X}~\cite{2022JATIS...8d4003C} 
and \textit{newASTROGAM}~\cite{Berge:2025ICRC}
sensitivities for high-latitude FBs and Loop I after 3 years of observations are shown by blue dash-dot-dotted and cyan long-dashed lines respectively.
}
\label{fig:FB_LoopI}
\end{figure}

Extrapolations to MeV energies of leptonic and hadronic models of \gray emission measured by the \Fermi~LAT~\cite{2014ApJ...793...64A_FB_Fermi} for the high-latitude FBs and Loop I are presented in~\autoref{fig:FB_LoopI}.
Both for the FBs and Loop I we exclude the Galactic plane within $|b| < 10^\circ$.
The corresponding CR proton and electron populations are modeled by power-law with exponential cutoff functions.
The parameters of the models are presented in \autoref{tab:FB_pars}.
The upper left plot shows models of the FBs assuming a magnetic field of 10\,$\mu$G, which is similar to the magnetic field in the models that can explain both the \gray and microwave haze emissions~\cite{2014ApJ...793...64A_FB_Fermi}.
For comparison, we also show the models with a much smaller magnetic field of 2\,$\mu$G.
The dashed purple line shows the leptonic model of the \gray emission.
In the leptonic model we do not take the energy loss into account, i.e., we model directly the spectrum of CR electrons inside the FBs.
The band shows the 68\% containment of the models taking into account statistical and systematic (added in quadrature) uncertainties of the measured FB spectrum~\cite{2014ApJ...793...64A_FB_Fermi}.
In the hadronic model (solid blue line) we take into account the \gray emission from the primary interactions of the CR protons with the ISG (orange dash dotted) and the IC (green sparse dash-dotted line) and bremsstrahlung (red dotted line) emission of the secondary leptons interacting with the ISRF and ISG, respectively.
For the ISRF we use a volume averaged model from \textsc{Galprop}~\cite{2008ApJ...682..400P, 2011CoPhC.182.1156V, 2017ApJ...846...67P} based on Ref.~\cite{1998ApJ...492..495F}.
We assume an ISG density of 0.01\,cm$^{-3}$~\cite{2007A&A...467..611F_Ferriere}.
With the 10\,$\mu$G magnetic field and such an ISG density, the dominant energy loss for the secondary leptons is synchrotron radiation.
As a result, one should expect the characteristic drop in the total \gray emission in the hadronic model below about 100 MeV (due to the mass of the $\pi^0$ meson).
In the leptonic model, on the other hand, one can expect approximately a power-law extrapolation of the emission below 100 MeV.
Consequently, the expected \gray emission between about 1 MeV and 100 MeV is lower in the hadronic model compared to the leptonic one.
The derived predictions are generally consistent with previous estimates of the \gray emission from the FBs at MeV energies~\cite{2018JHEAp..19....1D, Negro:2021urm}.

We show the expected sensitivity of \textit{COSI} after 2 years of observations~\cite{2023arXiv230812362T}, \textit{AMEGO-X} after 3 years~\cite{2020SPIE11444E..31K}, and \textit{newASTROGAM} after 3 years~\cite{Berge:2025ICRC} by yellow dashed, blue dash-dot-doted, and cyan long dashed lines respectively.
The sensitivity for intensity of emission from an extended source is estimated from the point source (PS) flux sensitivity as follows.
We assume that the PS flux sensitivity is dominated by a solid angle $\Omega_{\rm PS} = \pi R^2$, where we use either 68\% containment radius or the half maximum radius depending on the availability of the corresponding radii in the literature.
We also assume that the background is proportional to the solid angle
and that the statistical fluctuations in the background are proportional to the square root of the background, i.e., that the fluctuations satisfy Poisson statistics.
Thus, we estimate the flux sensitivity for an extended source subtending a solid angle 
$\Omega_{\rm ext}$ as $F_{\rm ext} = F_{\rm PS} \sqrt{\Omega_{\rm ext} / \Omega_{\rm PS}}$.
The corresponding sensitivity for intensity of emission is then $I_{\rm ext} = F_{\rm ext} / \Omega_{\rm ext} = F_{\rm PS} / \sqrt{\Omega_{\rm ext} \Omega_{\rm PS}}$.
We note that since 
$F_{\rm PS} \propto \sqrt{\Omega_{\rm PS}}$
in the regime where the sensitivity is dominated by the background, $I_{\rm ext}$ does not depend on $\Omega_{\rm PS}$ as long as the size of the extended source is much larger than the PSF radius.
%The sensitivities are obtained by dividing the point-source sensitivity with the square root of the solid angles $\sqrt{\Omega_{PS} \Omega_{\rm ROI}}$, where $\Omega_{\rm ROI} = 1$~sr is the approximate solid angle of the high-latitude FBs and $\Omega_{PS}$ is an effective area of emission from a point-like source (PS). 
For \textit{newASTROGAM} and \textit{AMEGO-X} we take $\Omega_{PS} = \pi R_{68\%}^2$.
For \textit{newASTROGAM}, we use a characteristic value of  the 68\% containment radius at 50 MeV of $4^\circ$, which is slightly larger than the corresponding radius for the 
\textit{e-ASTROGAM} configuration~\cite{2018JHEAp..19....1D}.
For \textit{AMEGO-X}, we use the 68\% containment radius of
$5^\circ$ at 50 MeV~\cite{2020SPIE11444E..31K}. For \textit{COSI}, we use the half-width half-maximum radius of $2^\circ$~\cite{2023arXiv230812362T}.
We find that for the 10\,$\mu$G magnetic field, the difference between the leptonic and hadronic models can be detected by the \textit{newASTROGAM} and \textit{AMEGO-X} experiments in the energy range between about 20 and 100 MeV.
This conclusion strongly depends on the assumption about the magnetic field.
For a much smaller magnetic field of, e.g., 2\,$\mu$G the hadronic model is practically indistinguishable from the leptonic model down to hundreds of keV, where the difference is at a sub-percent level compared to the isotropic diffuse background (cf.~\autoref{fig:egb_measurements}).
Thus, a presence of a break in the spectrum below about 100 MeV is a strong support for the hadronic origin of the \gray emission. 
In the absence of a break below 100 MeV, an independent assessment of the magnetic field is necessary: for the magnetic fields on the order of 10\,$\mu$G or above the leptonic model is preferred, while for much smaller magnetic fields than 10\,$\mu$G both leptonic and hadronic models are possible.

\begin{table}[]
\centering
\small{
\begin{tabular}{l|c|c|c}
\hline 
Model & $B$ ($\mu$G) & Index & $E_{cut}$ (GeV) \\ 
\hline 
FB hadronic & 2, 10 & $2.06 \pm 0.04$ & ${2200}^{+1100}_{-750}$ \\
FB leptonic & 2, 10 & $2.06 \pm 0.18$ & ${970}^{+320}_{-240}$ \\\hline 
Loop I hadronic & 2 & $2.60 \pm 0.15$ & ${350}^{+980}_{-260}$ \\
Loop I hadronic & 10 & $2.65 \pm 0.13$ & ${460}^{+1700}_{-360}$ \\
Loop I leptonic & 2, 10 & $2.39 \pm 0.28$ & ${21}^{+11}_{-7}$ \\
\hline 
\end{tabular}
}
\caption{Parameters of the best-fit leptonic and hadronic models of the FBs and Loop I at $|b| > 10^\circ$ assuming two different values for the magnetic field $B = 2$, 10 $\mu$G. The magnetic field only affects the secondary population of leptons in the hadronic model.}
\label{tab:FB_pars}
\end{table}

The corresponding models for Loop I are shown in \autoref{fig:FB_LoopI} on the right panels.
The 10\,$\mu$G and 2\,$\mu$G cases are on the upper and lower panels, respectively.
There is a break in the spectrum of the hadronic model below 100 MeV both for low and for high magnetic fields.
The main reason is that the spectrum of the primary protons is softer in the Loop I case compared to the FBs (cf.~\autoref{tab:FB_pars}), which results in a much smaller injected energy density of the secondary leptons in the Loop I case.
As a result, even for the 2\,$\mu$G magnetic field, the hadronic model is signifiantly below the extrapolation of the leptonic model at energies $E < 100$ MeV.
Although leptonic models of \gray emission from Loop I are generally preferred, especially in the GC outflow models of Loop I, there is still a possibility that the \gray emission has hadronic origin~\citep{2018Galax...6...27K}, e.g., in the superbubble scenario~\citep{2007ApJ...664..349W, 2018Galax...6...62S}.
In this case, the situation is similar to supernova remnants, where the protons and nuclei 
can be responsible for the majority of the \gray production, while electrons are responsible for the synchrotron radio emission.
An observation of a drop in the \gray emission below 100 MeV would be a signature of a hadronic origin of \gray emission from Loop I and an argument in favor of their local origin, although hadronic emission in Loop I in the GC outflow is also possible, e.g.,~\cite{2018Galax...6...27K}.
A flat SED below 100 MeV, on the other hand, would point to a leptonic origin of the \gray emission.
In this case, both local and GC outflow models of Loop I are possible.
Apart from Loop I, there are several other loops and spurs observed in radio data~\cite{1971A&A....14..252B, 2015MNRAS.452..656V, 2016A&A...594A..25P}.
At least one of these loops (Loop IV) is also detected in the \gray data~\cite{2021ApJ...917...30J}.
Comparison of Loop I with the other loops may shed light on whether Loop I is a special feature or all loops have similar origin.

\section{Extragalactic diffuse background}
\label{sec:extragalactic}
%Past measurements, current knowledge about origin, prospects for future missions, potentials for constraints on new physics 

\begin{figure}[thp]
\centering
\includegraphics[width=0.75\textwidth]{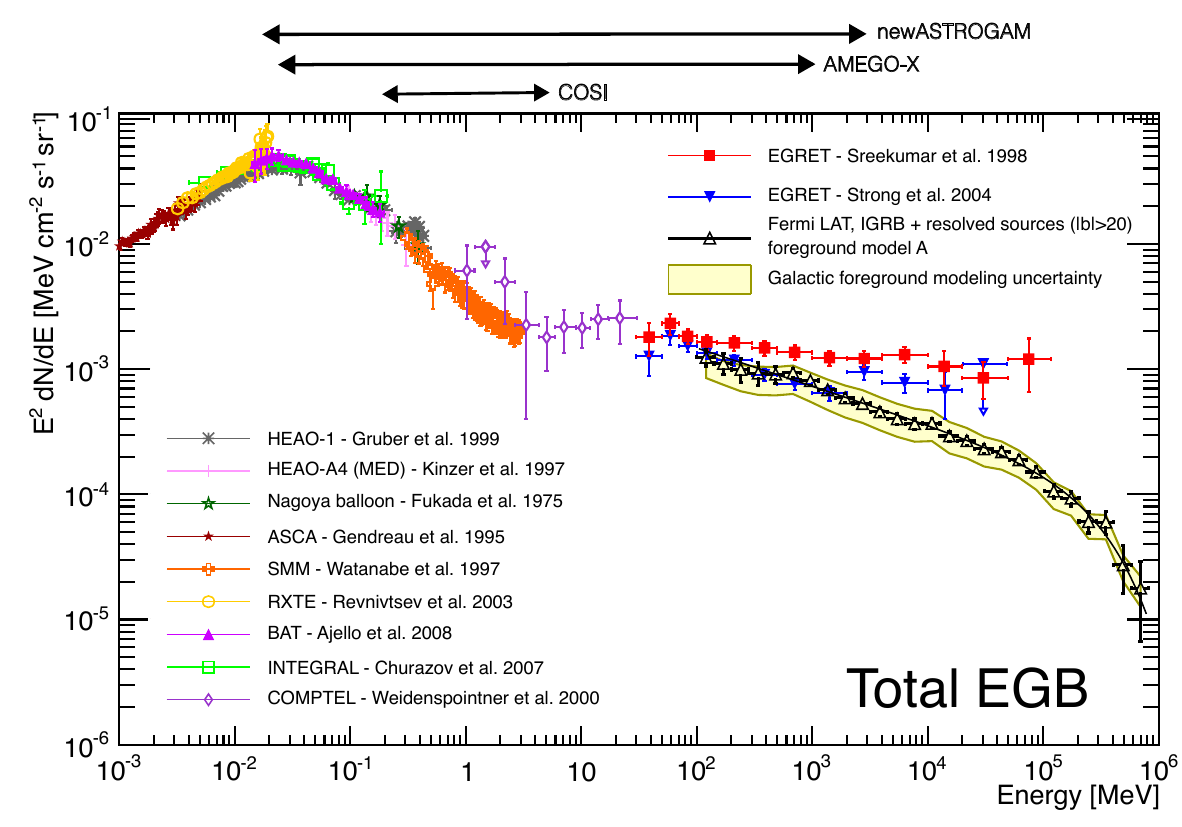}
\caption{Measurements of the extragalactic X-ray and \gray background from 1~keV to 820~GeV. The energy ranges of the upcoming \textit{COSI} satellite and other proposed future \gray space missions are indicted by the arrows above the figure. Figure adapted from~\cite{2015ApJ...799...86A}.}
\label{fig:egb_measurements}
\end{figure}

The first measurement of the MeV extragalactic background reaches all the way back to the \textit{Apollo} missions~\cite{1977ApJ...212..925T}, which detected a diffuse emission component in the 0.3-10 MeV range with a high degree of isotropy. 
The existence of this cosmic radiation background was confirmed by \textit{HEAO-1}~\cite{1997ApJ...475..361K}, \textit{SMM}~\cite{1999HEAD....4.0601W}, \textit{COMPTEL}~\cite{2000AIPC..510..467W}, and other measurements in the MeV band. 
It was extended to GeV energies by \textit{EGRET}~\cite{1998ApJ...494..523S}, and the most recent measurement above 100~MeV up to a maximum energy of 820~GeV was provided by the \Fermi LAT~\cite{2015ApJ...799...86A}.
\autoref{fig:egb_measurements} provides a selection of measurements of the extragalactic X-ray and \gray background from 1~keV to 1~TeV.

A substantial fraction -- if not all -- of the IGRB emission is customarily attributed to unresolved sources from the various extragalactic \gray source populations. In the MeV to GeV energy range, emission from star-forming galaxies (SFGs)~\cite{2014ApJ...786...40L}, jetted and non-jetted active galactic nuclei (AGN), such as Seyfert galaxies \cite{2008ApJ...672L...5I}, radio galaxies~\cite{2011ApJ...733...66I}, and blazars~\cite{2009ApJ...699..603A}, as well as the \gray emission from nuclear decays in cosmic supernovae, in particular SN Ia~\cite{1975ApJ...198..241C,2016ApJ...820..142R}, all contribute to the IGRB. 

\begin{figure}[thp]
    \centering
    \hspace{0.7cm}\includegraphics[width=0.825\textwidth]{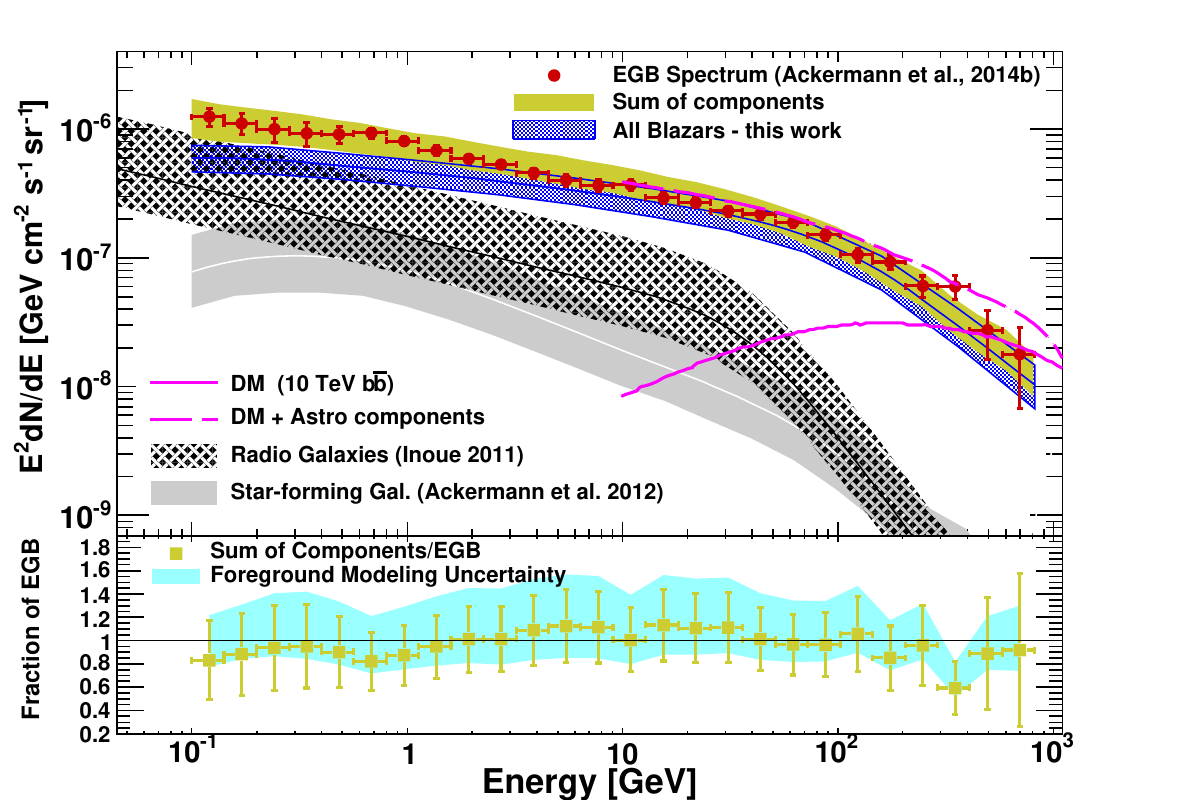}
    \caption{Overview of the contributions of various source populations to the total EGB. The shaded areas represent the uncertainties in the contributions according to the calculations in the respective publications. The magenta line demonstrates how a potential \gray signal from dark matter annihilation would lead to features in the spectrum. Non-observation of such features allows to constrain the dark matter annihilation cross section. Figure taken from \cite{2015ApJ...800L..27A}. }
    \label{fig:components}
    \end{figure}

While non-jetted AGN dominate the IGRB in the soft and hard X-ray range~\cite{2004NuPhS.132...86H}, blazar emission is the dominant source of the IGRB above few tens of~GeV~\cite{2016PhRvL.116o1105A}. The fractional contributions of the various populations in the MeV and low GeV range is still under debate and one of the prime open questions of \gray astronomy. 
The energy range between few hundred keV and few tens of MeV might be dominated by Seyfert galaxies~\cite{2019ApJ...871..240A} or blazars~\cite{2009ApJ...699..603A}, with unknown contributions from supernovae and jetted AGN~\cite{2016ApJ...820..142R}. Apparent breaks in the IGRB spectrum at few MeV and few tens of MeV could be interpreted as an indication for the relevance of multiple source populations in this energy range. Existing measurements by \textit{SMM} and \textit{COMPTEL} do not constrain the IGRB spectrum enough to identify individual contributions based on their characteristic spectral shapes. However, the upcoming \textit{COSI} satellite mission is expected to improve the precision of the spectral measurement significantly in the energy range between few hundred keV and few MeV.
Starforming galaxies are expected to contribute significantly to the IGRB above few tens of MeV with widely varying predictions (e.g.,~\cite{2014JCAP...09..043T,2017PhRvD..96h3001L,2020ApJ...894...88A}). A similar large range of predictions exists for the contribution of radio galaxies to the IGRB (e.g.,~\cite{2011ApJ...733...66I,2014ApJ...780..161D,2016JCAP...08..019H}). \autoref{fig:components} shows a schematic overview of the contributions of various source populations to the EGB, based on calculations in~\cite{2011ApJ...733...66I,2012ApJ...755..164A,2015ApJ...800L..27A} and their quoted uncertainties.

Disentangling the various contributions depends on modeling of the \gray luminosity functions of the various populations and their spectral shapes in the energy range of interest. 
The obtained intensity can then be compared to the intensity of the total EGB. For most populations only few sources were detected with past and current instruments in the MeV range, too few to  derive their luminosity functions directly. 
Consequently, observed or expected correlations between \gray luminosities and luminosities in other bands of the electromagnetic spectrum are used to estimate the \gray luminosity functions, leading to a substantial statistical and systematic uncertainty (see, e.g.,~\cite{2011ApJ...733...66I,2012ApJ...755..164A}). 
Similarly, the spectral shape of sources is often interpolated from X rays and high-energy \grays due to a lack of data in the MeV range.

Future \gray telescopes with improved sensitivity and angular resolution in the MeV range, will be able to advance our understanding of the IGRB in two ways. 
First, they will provide a more precise measurement of the IGRB and its spectral shape, due to their higher collection area and larger field-of-view. 
Second, they will allow to resolve a larger fraction of the total EGB into its individual source populations, providing a more direct measurement of their luminosity functions and spectral shapes. 

The IGRB and total EGB are also of fundamental importance for searches for BSM physics. Since our Universe is transparent to \grays, in particular in the MeV and low-GeV regimes, the \gray background provides strict constraints on the cumulative \gray emission from the observable Universe. 
This can be used to test, e.g., models of dark matter annihilation or decay~\cite{2015JCAP...09..008F}, or constrain the density of primordial black holes~\cite{2010PhRvD..81j4019C}. Precise measurements and a detailed understanding of the IGRB are crucial to unleash the full potential of these searches that need to disentangle astrophysical contributions from unresolved sources and potential \gray emission from non-standard model particles and/or primordial black holes.

In multi-messenger astronomy, the EGB at MeV and GeV energies is a key observable due to is intrinsic connection to the extragalactic neutrino background (ENB) at TeV and PeV energies~\cite{2020PhRvL.125l1104A,2022ApJ...928...50A} measured by the \textit{IceCube} neutrino telescope~\cite{2017JInst..12P3012A}. 
The ENB, much like the EGB, is considered to arise from unresolved neutrino sources. 

The standard neutrino production process in astrophysical sources is the decay of charged pions produced in the interactions of nucleons with gas and photon targets in the local source environments. 
This process inevitably leads to the simultaneous production of high-energy photons (with similar energies as the neutrinos) from the decay of neutral pions generated in the same interactions. In contrast to neutrinos, TeV and PeV photons cannot propagate over cosmological distances, but interact with the extragalactic background light (EBL, e.g.,~\cite{2017A&A...603A..34F}). 
Repeated pair production and IC interactions result in a cascading of the initial photons to GeV energies, at which the Universe is transparent for \grays even at cosmological distances.
This cascading in the EBL links the TeV/PeV neutrino background to the GeV \gray background and has been used to constrain properties of the neutrino sources and the neutrino spectrum below the energy range observable by \textit{IceCube}~\cite{2016PhRvD..94j3006M}.

Even more extreme cascading processes can alter the \gray spectra within the source environments close to the interaction region, e.g., within the photon targets that are producing the charged and neutral pions. 
In this case, intense UV to X-ray target photon fields would lead to a cascading even of GeV photons, which would finally escape at MeV energies. Such sources have been named ``hidden'' CR accelerators in~\cite{2016PhRvL.116g1101M} since they might be invisible in the GeV band, and unobserved in the MeV band, where the sensitivity of current instruments trails sensitivities achieved for GeV \grays. 
Remarkably, the first observation of a neutrino source, the Seyfert galaxy NGC~1068, shows a \gray luminosity in the GeV band that is more than an order of magnitude lower than its neutrino luminosity~\cite{2022Sci...378..538I}, pointing to the described cascading in this source. 

Models, such as~\cite{2022ApJ...941L..17M}, of the broad-band emission of NGC~1068 consequently predict a strong emission in the MeV band, where the source becomes transparent to photons. 
The MeV \gray background holds the imprint of all such sources in the Universe and can be used to constrain the population properties of ``hidden'' accelerators; in particular, in conjunction with the improved sensitivity of future MeV instruments and future neutrino telescopes that will allow to also individually detect the brightest of theses sources in \grays and neutrinos.

The measurement of the isotropic \gray background is challenging for various reasons. Strong Galactic foregrounds lead to large systematic uncertainties in the determination of the IGRB in the GeV band (see \autoref{fig:egb_measurements}). 
In the MeV band, the signal-to-noise between isotropic and Galactic gamma-ray emission improves significantly, however, new challenges arise: Secondary charged particles and \grays produced by CRs in the Earth's atmosphere can mimic the isotropic \gray background. 
While a part of the background can be suppressed by active veto systems, some background remains irreducible, such as \grays produced in secondary positron annihilation in passive material that inevitably surrounds any active shielding. 

This secondary cosmic-ray contamination dominates the instrumental background for the IGRB measurement at few hundred MeV~\cite{2012ApJS..203....4A}. 
In addition, the fluxes of secondary charged particles with energies below few GeV in low-Earth orbit (LEO) are poorly measured~\cite{2004ApJ...614.1113M,2019ExA....47..273C}, leading to large systematic uncertainties in modeling the expected background. 
Future mission concepts should keep these challenges in mind when optimizing the design of the instrument and the orbit, e.g., by minimizing passive material outside of active veto systems and/or devising strategies to measure the secondary charged particle flux in-situ.

At tens of MeV, the Earth's albedo \gray emission and unresolved Galactic sources can become a significant background for the IGRB measurement. The quickly deteriorating PSF in pair-conversion telescopes~\cite{2012ApJS..203....4A,2018JHEAp..19....1D} at these energies leads to source confusion and makes it increasingly hard to reject the Earth emission. 
Below few MeV, a new background arises from the activation of spacecraft material due to CRs and trapped particles in orbit, leading to a significant increase in the instrumental background over time~\cite{2001A&A...368..347W}. Careful monitoring and modeling of this background is required to subtract it from the measured data. Again, choice of orbit, in particular avoiding the trapped particle populations in the South Atlantic Anomaly, and the design of the instrument can help to minimize this background.

% What else?

\section{Conclusions}
\label{sec:conclusions}
Continuum diffuse \gray emission has been observed and studied since the early days of spaceflight that made their detection possible. Their main contributions, the DGE and the EGB are very different in nature. However, both contain essential clues for fundamental questions of physics and astronomy, such as the propagation of CRs in our Galaxy, or the nature of the non-baryonic matter in our Universe. 
Closely connected are large extended sources in the Milky Way, such as the FBs and Loop I, which have been observed from radio to GeV energies and provide unique information about the history of CR injection and high-power outflows in the Galaxy.

In this article we have demonstrated that a precise measurement of the spectrum and spatial distribution of the diffuse \gray emission in the MeV range is crucial for our ability to understand the origin of the emission.
In the case of the DGE, this will allow us to separate the components that arise from interactions of CRs with the ISG and ISRF. 
It also will give us an independent handle on the LIS spectrum of CR electrons in an energy range where local observations are dominated by solar modulation effects, but which is critical for testing different CR propagation scenarios. 

In the case of the EGB, a precise measurement of its spectrum will help us to identify and quantify the contributions of the various source populations in the \gray sky, and constrain potential additional contributions, e.g., from the annihilation of dark matter particles or the evaporation of primordial black holes. In a multi-messenger approach it will also connect neutrino and \gray observations, and help in understanding the properties of high-energy neutrino sources. 

In the case of extended sources, such as FBs and Loop I, high sensitivity in the MeV to hundred MeV range will allow one to distinguish in many scenarios leptonic and hadronic models of \gray production, which in turn will put constraints on the origin of these structures. In particular, leptonic models are favored in the AGN scenario of FB formation, while hadronic emission is generally preferred in scenarios involving star formation and supernovae.

The upcoming \textit{COSI} satellite mission, covering the energy range from few hundred keV to few MeV with unprecedented sensitivity, is a crucial first step towards achieving the described advancements in our understanding of the origin of diffuse \gray emission and large extended structures in the \gray sky. Ultimately, it will, however, require a new generation of MeV telescopes, such as \textit{newASTROGAM} or \textit{AMEGO-X}, which are one to two orders of magnitude more sensitive than past and current instruments and cover a wide energy range from sub-MeV to low-GeV energies.

\section*{Declarations}
All authors certify that they have no affiliations with or involvement in any organization or entity with any financial interest or non-financial interest in the subject matter or materials discussed in this manuscript.

\bibliography{MeV_diffuse}% common bib file

\end{document}